\documentclass[12pt,preprint]{aastex}

\def\<<{{\ll}}
\def\>>{{\gg}}

\def\spose#1{\hbox to 0pt{#1\hss}}
\def\ltwig{\mathrel{\spose{\lower 3pt\hbox{$\mathchar"218$}}
     \raise 2.0pt\hbox{$\mathchar"13C$}}}
\def\gtwig{\mathrel{\spose{\lower 3pt\hbox{$\mathchar"218$}}
     \raise 2.0pt\hbox{$\mathchar"13E$}}}
\def\+/-{{\pm}}
\def\=={{\equiv}}
\def\Rstar{R_{\ast}}

\def\vth{v_{th}}

\def\vinf{v_\infty}

\newcommand{\beq}{\begin{equation}}
\newcommand{\eeq}{\end{equation}}

\begin{document}
    
\title{Dynamical Simulations of Magnetically Channeled Line-Driven 
Stellar Winds: 
I. Isothermal, Nonrotating, Radially Driven Flow
}
              
\author{Asif ud-Doula and Stanley P. Owocki}
\affil{Bartol Research Institute,
       University of Delaware, 
       Newark, DE 19716}   


\begin{abstract}

We present numerical magnetohydrodynamic (MHD) simulations of
the effect of stellar dipole magnetic fields on line-driven
wind outflows from hot, luminous stars.
Unlike previous fixed-field analyses, the simulations here take
full account of the dynamical competition between field and flow, 
and thus apply to a full range of magnetic field strength,
and within both closed and open magnetic topologies.
A key result is that the overall degree to which
the wind is influenced by the field depends largely on a single, 
dimensionless, `wind magnetic confinement parameter', $\eta_{\ast}$ 
($ = B_{eq}^2 R_\ast^2/{\dot M} v_\infty$), which characterizes the 
ratio between magnetic field energy density and kinetic energy
density of the wind. 
For weak confinement $\eta_{\ast} \le 1$, the field is fully
opened by the wind outflow, but nonetheless for confinements as small
as $\eta_{\ast}=1/10$ can have a significant back-influence in enhancing
the density and reducing the flow speed near the magnetic equator.
For stronger confinement $\eta_{\ast} > 1$, the magnetic field remains
closed over a limited range of latitude and height about the equatorial
surface, but eventually is opened into a nearly radial configuration
at large radii.
Within closed loops, the flow is channeled toward loop tops into shock
collisions that are strong enough to produce hard X-rays,
with the stagnated material then pulled by gravity back onto the star 
in quite complex and variable inflow patterns.
Within open field flow, the equatorial channeling leads to oblique
shocks that are again strong enough to produce X-rays, and also lead to a 
thin, dense, slowly outflowing `disk' at the magnetic equator.
The polar flow is characterized by a faster-than-radial expansion
that is more gradual than anticipated in previous 1D flow-tube analyses,
and leads to a much more modest increase in terminal speed ($ < 30\%$),
consistent with observational constraints.
Overall, the results here provide a dynamical groundwork for 
interpreting many types of observations -- e.g., UV line profile 
variability; red-shifted absorption or emission features; enhanced
density-squared emission; X-ray emission -- that might be associated
with perturbation of hot-star winds by surface magnetic fields.
\end{abstract}


\section{Introduction}                              
\label{sec:introduction}

Hot, luminous, OB-type stars have strong stellar winds, with 
asymptotic flow speeds up to $v_\infty \sim 3000$~km/s and mass
loss rates up to ${\dot M} \sim 10^{-5} M_\odot$~/yr.
These general properties are well-explained by modern extensions
(e.g. Pauldrach, Puls, and Kudritzki 1986)  of the basic formalism
developed by  
Castor, Abbott, and Klein (1975; hereafter CAK)
 for wind driving by scattering of the star's continuum radiation in a 
large ensemble of spectral lines.

However there is also extensive evidence that such winds are not the
steady, smooth outflows envisioned in these spherically symmetric,
time-independent, CAK-type models,
but instead have extensive structure and variability on a range of
spatial and temporal scales.
Relatively small-scale, stochastic structure -- e.g. as evidenced by
often quite constant soft X-ray emission (Long \& White 1980), or by
UV lines with extended black troughs understood to be a signature of a
nonmonotonic velocity field (Lucy 1982) -- seems most likely
a natural result of the strong, intrinsic instability of the
line-driving mechanism itself (Owocki 1994; Feldmeier 1995).
But larger-scale structure -- e.g. as evidence by explicit UV line
profile variability in even low signal-to-noise IUE spectra 
(Kaper et al. 1996; Howarth \& Smith 1995) --
seems instead likely to be the consequence of wind perturbation
by processes occurring in the underlying star.
For example, the photospheric spectra of many hot stars show evidence
of radial and/or non-radial pulsation, and in a few cases there is
evidence linking this with observed variability in UV wind lines
(Telting, Aerts, \& Mathias 1997; Mathias et al. 2001).

An alternate scenario -- one explored through dynamical
simulations here -- is that, in at least some hot stars, surface 
magnetic fields could perturb, and perhaps even channel, the wind 
outflow, leading to rotational modulation of wind structure that
is diagnosed in UV line profiles, and perhaps even to magnetically
confined wind-shocks with velocities sufficient to produce the
relatively hard X-ray emission seen in some hot-stars.

The sun provides a vivid example of how a stellar wind can be 
substantially influenced by a surface magnetic field.
Both white-light and X-ray pictures show the solar corona to be highly
structured, with dense loops where the magnetic field confines the 
coronal gas, and low-density coronal holes where the more radial 
magnetic field allows a high-speed, pressure-driven, coronal outflow
(Zirker 1977).
In a seminal paper, Pneuman and Kopp (1971) provided the first
magnetohydrodynamical (MHD) model of this competition between magnetic
confinement and coronal expansion.
Using an iterative scheme to solve the relevant partial differential
equations for field and flow,
they showed that this competition leads naturally to the commonly 
observed `helmet' streamer 
configuration, for which the field above closed magnetic loops is extended 
radially outward by the wind outflow.
Nowadays such MHD processes can be readily modelled using 
time-dependent MHD simulation codes, such as the Versatile Advection 
Code (Keppens and Goedbloed 1999), or the publicly available ZEUS codes
(Stone and Norman 1992). 
Here we apply the latter to study MHD processes 
within {\it line-driven} stellar winds that have many characteristics
quite distinct from the {\it pressure-driven} solar wind.

For the solar wind, the acceleration to supersonic speeds
can take several solar radii; as such, magnetic loops that 
typically close within a solar radius or so can generally
maintain the coronal gas in a nearly hydrostatic configuration.
As we show below, in the more rapid line-acceleration of hot-stars 
winds, strong magnetic confinement typically channels an already supersonic
outflow, often leading to strong shocks where material originating
from different footpoints is forced to collide, with compressed material 
generally falling back to the star in quite complex and chaotic patterns.
In the solar wind, the very low mass-loss-rate, and thus the low gas density 
and pressure, mean that only a modest magnetic field strength, of 
order of a Gauss, is sufficient to cause significant confinement and
channeling of the coronal expansion.
In hot-star winds,
magnetic confinement or channeling generally requires a much
stronger magnetic field, on the order of hundreds of Gauss.
As such, a key issue underlying the study here regards the theoretical
prospects and observational evidence for hot-star magnetic fields of this 
magnitude.

In the sun and other cool stars, magnetic fields are understood to be
generated through a dynamo mechanism, in which coriolis forces 
associated with stellar rotation deflect convective motions in the
hydrogen and helium recombination zones.
In hot stars, hydrogen remains fully ionized even through the 
atmosphere, and so, lacking the strong convection zones associated
with hydrogren recombination, such stars have traditionally been
considered not to have strong, dynamo-generated magnetic fields.
However, considering the generally quite rapid rotation of most
hot stars, dynamo-generation may still be possible, e.g. within thin,
weaker, near-surface convection zones associated with 
recombination of fully ionized helium.

Moreover, the interior, energy-generation cores of such massive
stars are thought to have strong convection, and recently 
Cassinelli and Macgregor (2000; see also Charbonneau and MacGregor 2001) 
have proposed that dynamo-generated magnetic flux tubes from this 
interior could become buoyant, and thus drive an upward diffusion to the 
surface over a time-scale of a few million years. 
Such a model would predict surface appearance of magnetic fields
in hot-stars that have evolved somewhat from the zero-age-main sequence.
Alternatively, magnetic fields could form from an early, convective 
phase during the star's initial formation, or perhaps even arise  through 
compression of interstellar magnetic flux during the initial stellar
collapse.
Such primordial models would thus predict magnetic fields to be strongest in the
youngest stars, with then some gradual decay as the star evolves.

In recent years there has been considerable effort to develop new
techniques (e.g., based on the Hanle effect; 
Ignace, Nordsieck, \& Cassinelli 1997;
Ignace, Cassinelli, \& Nordsieck 1999) to observationally detect stellar magnetic fields.
But the most direct and well-demonstrated method is
through the Zeeman splitting and associated circular polarization of
stellar photospheric absorption lines (Borra \& Landstreet 1980).  
This technique has been used extensively in direct measurement of
the quite strong magnetic fields that occur in the chemically
peculiar Ap and Bp stars (Babcock 1960; Borra et al. 1980; Bohlender 1993; 
Mathys 1995; Mathys et al. 1997).
For more `normal' (i.e., chemically non-peculiar) hot stars, the 
generally strong rotational line-broadening severely hinders the direct
spectropolarimetric detection of their generally much weaker fields, yielding 
instead mostly only upper limits, typically of order a few hundred Gauss.
This, coincidentally and quite tantalizingly, is similar to the level at which
magnetic fields can be expected to become dynamically significant for 
channeling the wind  outflow.

Recently, however, there have been first reports of positive field
detections in a few normal hot stars.
For the relatively slowly rotating star $\beta$~Cephei, 
Henrichs et al. (2000) and Donati et al. (2001) report a 
3-sigma detection of a ca. 400 G dipole field, with moreover a
rotational modulation suggesting the magnetic axis
is tilted to be nearly perpendicular to the stellar rotation.
There are also initial reports (Donati 2001) of a ca. 1000 G
dipole field in $\theta^1$~Ori~C, this time with the magnetic axis
tilted by about 45 degrees to the rotation.
For Ap and Bp stars, the most generally favored explanation for their
strong fields is that they may be primordial, and 
the relative youth of $\theta^1$~Ori~C also seems to suggest
a primordial model.
In contrast, the more evolved evolutionary status
of $\beta$~Cephei seems to favor the interior-eruption scenario.
\par
The focus of the present paper is to carry out magnetohydrodynamical
simulations of how such magnetic fields on the surface of hot stars 
can influence their radiatively driven stellar wind.
Our approach here represents a natural extension of the previous
studies by Babel \& Montmerle (1997a,b; hereafter BM97a,b), which effectively 
{\it prescribed} a fixed magnetic field geometry to channel the wind outflow.
For large magnetic loops, wind material from opposite footpoints
is accelerated to a substantial fraction of the wind terminal speed 
(i.e. $\sim 1000$~km/s) before the channeling toward the loop tops forces a 
collision with very strong shocks, thereby heating the gas to temperatures 
($10^7-10^8$~K) that are high enough to emit hard (few keV) X-rays.
This `magnetically confined wind shock' (MCWS) model was initially
used to explain X-ray emission from the Ap-Bp star IQ Aur
(BM97a), which has a quite strong magnetic field
($\sim 4$kG) and a rather weak wind (mass loss rate $\sim 10^{-10} 
M_{\odot}/$~yr), and thus can indeed be reasonably modeled within
the framework of prescribed magnetic field 
geometry.\footnote{However, note that even in this case the more-rapid 
radial decline in magnetic vs. wind energy density means that the wind
outflow eventually wins, drawing out portions of the surface field 
into a radial, open configuration. Such open-field regions can only 
be heuristically accounted for in the fixed-field modeling approach.}
Later, BM97b applied this model to explain
the periodic variation of X-ray emission of the O7 star
$\theta^1$~Ori~C, which has a much lower magnetic field ($\ltwig 1000$ G)
and significantly stronger wind (mass loss rate $\sim 10^{-7} 
M_{\odot}/$~yr), raising now the possibilty that the wind itself 
could influence the field geometry in a way that is not considered in the 
simple fixed-field approach.
\par
The simulation models here are based on an 
isothermal approximation of the complex energy balance,
and so can provide only a rough estimate of the level of shock 
heating and X-ray generation.
But a key advantage over previous approaches is that these models do
allow for such a fully dynamical competition between the field and flow.
A central result is that the overall effectiveness of magnetic field
in channeling the wind outflow can be well characterized in terms of
single `wind magnetic confinement parameter' $\eta_\ast$, defined in eqn. 
(\ref{wmcpdef}) below,  and related to the relative energy densities of field
and wind (\S 3).
The specifics of our numerical MHD method are described
in \S 2, while \S 4 details the results of a general parameter
study of hot-star winds with various degrees of magetic confinement.
Following a discussion (\S 5) of the implications of these results for 
modeling hot-star wind structure and variability, we finally conclude
(\S 6) with a summary  and outlook for future work.

\section{Numerical Method}

\subsection {Magnetoydrodynamic Equations}

Our general approach is to use the ZEUS-3D (Stone and Norman 1992)
numerical MHD code  to evolve a consistent dynamical 
solution for a line-driven stellar wind 
from a star with a dipolar surface field.
As described further below, the basic ZEUS-3D code was modified for the present 
study to include  radiative driving terms, and to allow for specification of the 
lower boundary conditions.
The code is designed to be easily adapted to run in a variety of
flow geometries (planar, cylindrical, spherical) in one, two, or
three dimensions.
Our implementation here uses spherical polar coordinates with
radius $r$, co-latitude $\theta$, and azimuth $\phi$ in a 
2D formulation, which assumes all quantites are constant
in azimuthal angle $\phi$, 
and that the azimuthal components of both field and flow vanish,
$B_\phi=v_\phi=0$.

The time-dependent equations to be numerically integrated thus 
include the conservation of mass,
\begin{equation}
 \frac{D \rho}{D t} + \rho \nabla \cdot {\bf v} = 0,
\label{masscon}
\end{equation}
and the equation of motion
\begin{equation}
 \rho \frac{D{\bf v}}{Dt} = - \nabla p +\frac{1}{4 \pi} (\nabla \times 
{\bf B}) \times {\bf B} - { GM (1-\Gamma ) {\hat {\bf r}} \over r^2 }
+ {\bf g}_{lines}
,
\label{eom}
\end{equation}
where $\rho$, $p$, and ${\bf v}$ are the mass density, gas pressure, 
and velocity of the fluid flow, and $D/Dt=\partial/\partial t+ 
{\bf v} \cdot \nabla$ is the advective time derivative.
The gravitational constant $G$ and stellar mass $M$ set the
radially directed (${\hat{\bf r}}$) gravitational acceleration, 
and $\Gamma \equiv \kappa_e L/(4 \pi G M c )$ is the Eddington 
parameter, which accounts for the acceleration due to 
scattering of the stellar luminosity $L$ by free electron opacity 
$\kappa_e$, with $c$ the speed of light.
The additional radiative acceleration due to {\it line} scattering,
${\bf g}_{lines}$, is discussed further below.
The magnetic field ${\bf B}$ is constrained to be divergence free
\begin{equation}
\nabla \cdot {\bf B}= 0 ,
\end{equation}
and, under our assumption of an idealized MHD flow with infinite 
conductivity (e.g. Priest \& Hood 1991),
its inductive generation is described by
\begin{equation}
 \frac{\partial {\bf B}}{\partial t}=\nabla \times
({\bf v} \times {\bf B}).
\end{equation}

The ZEUS-3D code can also include an explicit equation for conservation of energy,
but in the dense 
stellar winds considered here, the energy balance is dominated
by radiative processes that tend to keep the wind near the
stellar effective temperature $T_{eff}$ ( Drew 1989; Pauldrach 1987).
In this initial study, we assume an explicitly isothermal flow with
$T=T_{eff}$, which thus implies a constant sound speed $a=\sqrt{kT/m}$,
with $k$ Boltzmann's constant and $m$ the mean atomic weight of the 
gas.
The perfect gas law then gives the pressure as $p=\rho a^2$.
In such an isothermal model, even the locally strong compressive 
heating that occurs near shocks is assumed to be radiated away within a 
narrow, unresolved cooling layer 
(Castor 1987; Feldmeier et al. 1997; Cooper 1994).
We thus defer to future work the quantitative modeling of the 
possible EUV  and X-ray emission of any such shocks, although below 
(\S 5.3.4) we do use computed 
velocity jumps to provide rough estimates of the expected 
intensity and  hardness of such shock emissions. 
(See figure \ref{fig9}.)

\subsection{Spherically Symmetric Approximation for Radial Line-Force}

The radiative acceleration ${\bf g}_{lines}$ results from scattering of the 
stellar radiation in a large ensemble of spectral lines.
In these highly supersonic winds this can be modeled within the framework of the
Sobolev (1960) approximation as depending primarily on the {\it local} velocity gradient
averaged over the directions of the source radiation from the stellar disk.
For 1D nonrotating winds, the line-force-per-unit-mass can be written in the form
(cf. Abbott 1982; Gayley 1995)
\beq
\label{glines}
g_{lines}
=\frac{f}{(1-\alpha)} ~ \frac{\kappa_e L{\bar {Q}}}{4 \pi r^2 c }
\left (\frac{dv/dr}{\rho c \bar Q \kappa_e}\right )^{\alpha}
\eeq
where $\alpha$ is the CAK exponent, and 
$f$ is the (1D) finite disk correction factor, given by CAK eqn. (50). 
(See also Friend and Abbott 1986, and Pauldrach, Puls, and Kudritzki 1986.)
Here we choose to follow the Gayley (1995) line-distribution normalization 
$\bar Q$, which offers the advantages of being a dimensionless 
measure of line-opacity that is independent of the assumed ion thermal 
speed $\vth$, and with a nearly constant characteristic value of order 
${\bar Q} \sim 10^3$ for a wide range of ionization conditions (Gayley 
1995).
This normalization is related  to the ususal CAK parameter through $k= {\bar 
Q}^{1-\alpha} \, \left ( \vth/c \right)^{\alpha}/(1-\alpha)$. 

In the 2D wind models computed here, the line force 
(\ref{glines}) should in principle be modified to take account of gradients 
in other velocity components, such as might arise from,
e.g., latitudinal flow along magnetic loops.
Such latitudinal gradients can modify the form of the radial
finite-disk correction factor $f$, 
and can even lead to a non-zero
{\it latitudinal} component of the full vector line force.
In a rotating stellar wind, asymmetries in the velocity gradient 
between the approaching and receding stellar hemisphere can even lead
to a net {\it azimuthal} line force (Owocki, Cranmer, and Gayley 1996;
Gayley and Owocki 2000).
For simplicity, we defer study of such rotational and latitudinal
affects to future work, and thus apply here just
the radial, 1D form (\ref{glines}) for the line force,
within a 2D, axisymmetric model of a non-rotating stellar wind.

\subsection {Numerical Specifications}

Let us next describe some specifics of our numerical discretization and boundary 
conditions. In our implementation of the ZEUS MHD code,
flow variables are specifed on a fixed 2D spatial mesh
in radius and co-latitude,  $\{r_i,\theta_j\}$.
The mesh in radius is defined from an initial zone $i=1$, which has a 
left interface at $r_1  = \Rstar$, 
the star's surface radius, out to a maximum zone ($i=n_r=300$), which 
has a right interface at 
$r_{301}  = 6 \Rstar$.
Near the stellar base, where the flow gradients are 
steepest, the radial grid has an initially fine spacing with
$\Delta r_{1}=0.00026 \Rstar$, and then increases by 2\% per zone out to a 
maximum of $\Delta r_{299}=0.11 \Rstar$.

The mesh in co-latitude uses $n_\theta=100$ zones to span the two 
hemispheres from one pole, where the $j=1$ zone has a left interface at
$\theta_1=0 \degr$, to the other pole, where the $j=n_\theta=100$ zone 
has a right interface at $\theta_{101}=180 \degr$.
To facilitate resolution of compressed flow structure near the magnetic
equator at $\theta= 90 \degr$, the zone
spacing has a minimum of $\Delta \theta_{50}=0.29 \degr $ at the equator, and 
then increases by 5\% per zone toward each pole, where 
$\Delta  \theta_1=\Delta  \theta_{100}=5.5 \degr $.
Test runs with half the resolution in radius and/or latitude showed some 
correspondingly reduced detail in flow fine-structure, but overall
the results were qualitatively similar to those for the standard resolution.

Our operation uses the piecewise-linear-advection option within ZEUS
(VanLeer 1977), with time steps set to a factor 0.30 of the minimum 
MHD Courant time computed within the code (Courant et al. 1953).
Boundary conditions are implemented by specification of variables in
two phantom zones.
At both poles, these are set by simple {\it reflection} about the
boundary interface.
At the outer radius, the flow is invariably super-Afvenic outward, 
and so outer boundary conditions for all variables (i.e. density, and the
radial and latitutudinal components of both the velocity and magnetic 
field) are set by simple {\it extrapolation}
assuming constant gradients.

The  boundary conditions at the stellar surface are specified as follows.
In the two radial zones below $i=1$, we 
set the radial velocity $v_{r}$ by constant-slope extrapolation,
and 
fix the density at a value $\rho_o$
chosen to ensure subsonic base outflow for the characteristic
mass flux of a 1D, nonmagnetic CAK model, i.e.
$\rho_o \approx {\dot M}/(4 \pi \Rstar^2 a/5)$.
Detailed values for each model case are given in Table 1.
In our 2D magnetic models, these conditions allow the mass flux and the 
radial velocity to adjust to whatever is appropriate for
the local overlying flow
(Owocki, Castor, and Rybicki 1988).
In most zones, this corresponds to a subsonic wind outflow,  although inflow 
at up to the sound speed is also allowed.

Magnetic flux is introduced through the radial boundary as the radial
component of a dipole field $B_r (\Rstar) = B_o \cos (\theta)$, where
the assumed polar field $B_o$ is a fixed parameter for each model (see Table 1).
The latitudinal component of magnetic field, $B_\theta$, is set by constant slope
extrapolation.
The specification for the latitudinal velocity, $v_\theta$, differs 
for strong vs. weak field cases.
For strong fields (defined by the magnetic confinement parameter 
$\eta_{\ast} > 1$; see \S 3), we again use linear extrapolation.
(We also find similar results using $v_\theta=v_r B_\theta/B_r$.)
For weak fields ($\eta_\ast \le 1$), we simply set $v_\theta=0$.
Tests using each approach in the intermediate field strength case
($\eta_{\ast}=1$) gave similar overall results.

This time-dependent calculation also requires us to specify an {\it
initial condition} for each of these flow variables over the entire
spatial mesh at some starting time $t=0$.
The hydrodynamical flow variables $\rho$ and ${\bf v}$ are initialized 
to values for a spherically symmetric, steady, radial CAK 
wind, computed from relaxing a 1D, non-magnetic, wind simulation
to an asymptotic steady state.
The magnetic field is initilized to have a simple dipole form with
components 
$B_r=B_o (\Rstar / r)^3 \cos \theta$, 
$B_\theta= (B_o/2) (\Rstar / r)^3 \sin \theta$, 
and $B_\phi = 0$,
with $B_o$ the polar field strength at the stellar surface.
From this initial condition, the numerical model is then evolved 
forward in time to study the dynamical competition between the field and flow.
The results of such dynamical simulations are described in \S 4.

\section{Heuristic Scaling Analysis for Field vs. Flow Competition}

\subsection{The Wind Magnetic Confinement Parameter}

To provide a conceptual framework for interpreting our MHD simulations, 
let us first carry out a heuristic scaling analysis of the competition
between field and flow.
To begin, let us
define a characteristic parameter for the relative
effectiveness of the magnetic fields in confining and/or channeling the
wind outflow.
Specifically, 
consider the ratio between the energy densities of field vs. flow,
\begin{eqnarray}
\eta (r, \theta) &\equiv& \frac{B^2/8\pi}{\rho v^2/2}
\nonumber
\\
&\approx& \frac{B^2 r^2}{\dot{M}v}
\label{etadef}
\\
&=& \left [\frac{B_{\ast}^2 (\theta )
{R_{\ast}}^2}{\dot{M}v_{\infty}}\right ] 
\left [\frac{(r/R_{\ast})^{2-2q}}{1-R_{\ast}/r} \right ] \, ,
\nonumber
\end{eqnarray}
where the latitudinal variation of the surface field has the
dipole form given by
$B_\ast^2 (\theta) = B_o^2 ( \cos^2 \theta + \sin^2 \theta/4 )$.
In general, a magnetically channeled outflow will have a complex
flow geometry, but for convenience, the second equality in eqn. 
(\ref{etadef}) 
simply characterizes the wind  strength in terms of a spherically 
symmetric mass loss rate
$\dot{M}=4\pi r^2 \rho v$.
The third equality likewise characterizes 
the radial variation of outflow velocity
in terms of the phenomenological velocity law 
$v(r) =v_{\infty}(1-R_{\ast}/r)$,
with $v_{\infty}$ the wind terminal speed;
this equation furthermore models the magnetic field strength 
decline as a power-law in radius, $B(r) =B_{\ast}(R_{\ast}/r)^q$, 
where, e.g., for a simple dipole $q=3$.

With the spatial variations of this energy ratio thus isolated within the right 
square bracket, we see that the left square bracket represents
a dimensionless constant that characterizes the overall relative 
strength of field vs. wind.
Evaluating this in the region of the magnetic equator 
($\theta=90\degr$), 
where the tendency 
toward a radial wind outflow is in  most direct competition with the 
tendency for a horizontal orientation of the field, 
we can thus define a equatorial `wind magnetic confinement parameter',
\begin{eqnarray}
\eta_{\ast} &\equiv& 
\frac{B_\ast^2 (90\degr) {R_{\ast}}^2}
{\dot{M}v_{\infty}}
\nonumber
\\
&=& 0.4  \,  \frac{B_{100}^2 \, R_{12}^2}{\dot{M}_{-6} \, v_8}.
\label{wmcpdef}
\end{eqnarray}
where 
$\dot{M}_{-6} \equiv \dot{M}/(10^{-6}\, M_{\odot}$/yr), 
$B_{100} \equiv B_o/(100$~G),
$R_{12} \equiv R_{\ast}/(10^{12}$~cm), and 
$v_{8} \equiv v_{\infty}/(10^8$~cm/s). 
As these stellar and wind parameters are scaled to typical values for
an OB supergiant, e.g. $\zeta$ Pup, the last equality in eqn. (\ref{wmcpdef})
immediately suggests that for such winds, significant magnetic 
confinement or channeling should require fields of order $ \sim 
100$~G.
By contrast, in the case of the sun, the much weaker mass loss 
(${\dot  M}_\odot \sim 10^{-14}~M_{\odot}$/yr) means that even a 
much weaker global field ($B_{o} \sim 1$~G) is sufficient to yield 
$\eta_{\ast} \simeq 40$, implying a substantial magnetic 
confinement of the solar coronal expansion.
This is consistent with the observed large extent of magnetic loops
in optical, UV and X-ray images of the solar corona.

\subsection{Alfven Radius and Magnetic Closure Latitude}

The inverse of the energy density ratio defined in eqn. (\ref{etadef})
also represents the square of an Alfvenic Mach number $M_A \equiv v/v_A$,
where $v_A \equiv B/\sqrt{4 \pi \rho}$ is the Alfven speed.
We can thus estimate from eqns. (\ref{etadef}) and (\ref{wmcpdef})
that the Alfven radius $R_A (\theta) $ (at which $M_A (R_A) \equiv 1 $)
satisifies
\begin{equation}
    \left [ {R_A (\theta ) \over R_\ast } \right ]^{2q-2} -
    \left [ {R_A (\theta ) \over R_\ast } \right ]^{2q-3} = \eta_{\ast} 
    \left [ 4-3 \sin^2 (\theta ) \right ] \,  .
\label{radef}
\end{equation}

\begin{figure}
\begin{center}
\plotone{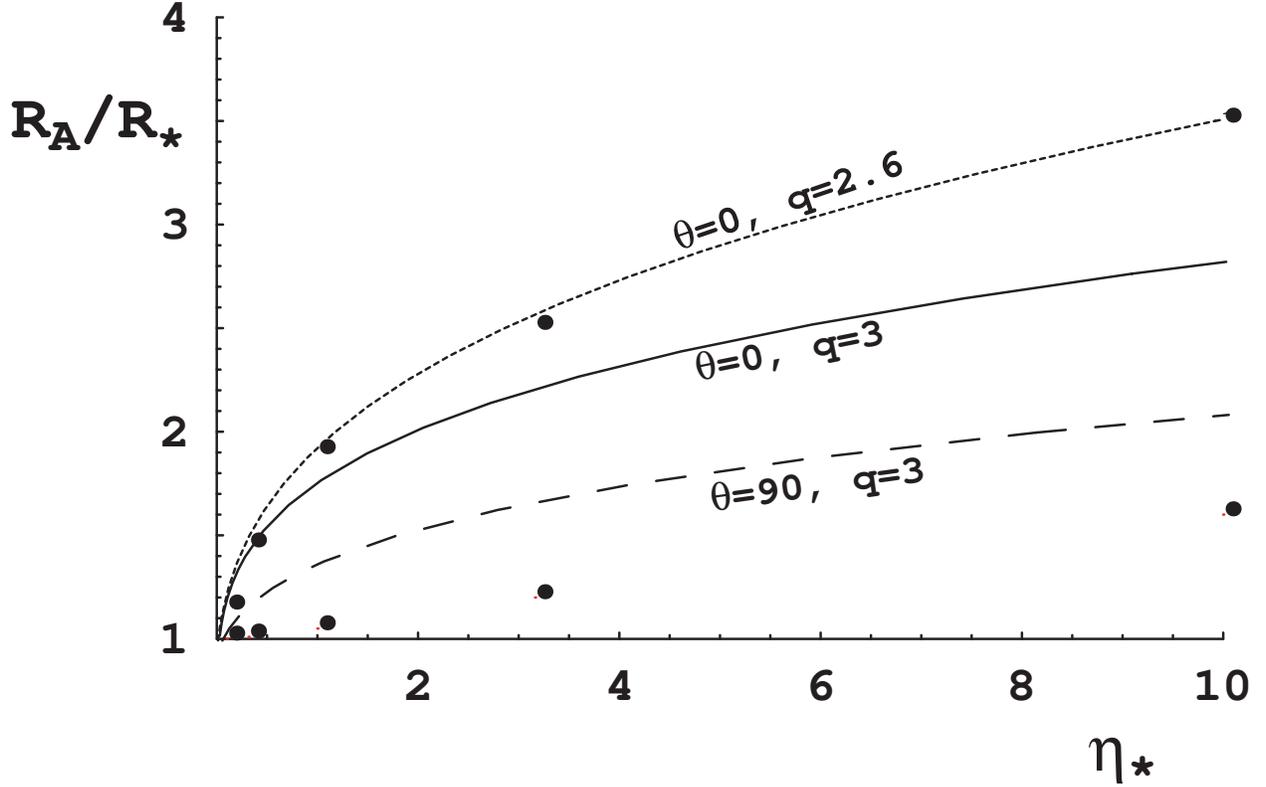}
\end{center}
\caption{
Variation Alfven radius $R_A$ with magnetic confinement parameter
$\eta_{\ast}$.
The points indicate MHD simulation model results above the pole (upper
set) or near the magnetic equator (lower set).
The lines show estimates from the magnetic field approximation of 
eqn. (\ref{radef}), applied at the  equator ($\theta=90 \degr$; dashed line), 
or at the pole ($\theta=0 \degr$), using radial decline power index for pure 
dipole ($q=3$; solid line) or modified by wind radial expansion 
($q=2.6$; dotted line).
}
\label{fig1}
\end{figure}

Figure \ref{fig1} plots solutions of this estimate of the polar 
($\theta=0$; solid curve)
and equatorial ($\theta=90 \degr $; dashed curve)
Alfven radius 
versus 
the confinement parameter $\eta_{\ast}$ for various values of the
magnetic exponent $q$. 
The circles compare the actual computed Alfven
radii near the magnetic poles (filled) and equator (open)
for our MHD simulations, as will be discussed further below 
(\S\S 4, 5.1).

This heuristic solution for the Alfven radius 
can be used (cf. BM97a) to estimate 
the maximum radius of a closed 
loop.\footnote{BM97a denote this as $L_A$, for the value of their 
`magnetic shell parameter' $L$ that intersects the Alfven radius.}
For an assumed dipole field, 
this loop has surface footpoints
at colatitudes $\theta_A$  satisfying
\begin{equation}
    \theta_A = \arcsin 
    \left [ \sqrt{ R_\ast \over R_A ( \theta=0 ) } \, \right ] \, ,
\label{tcdef}
\end{equation}
which thus give the latitudes $\pm |90-\theta_A| \degr$
bounding the  region of magnetic closure about the equator.

The discussion in \S 5.1 examines how well these heuristic scaling
arguments match the results of the full MHD simulations that
we now describe.

\section{MHD Simulation Results}
Let us now examine results of our MHD simulations for line-driven 
winds.
Our general approach is to study the nature of the wind outflow for
various assumed values of the wind magnetic confinement parameter 
$\eta_{\ast}$.
We first confirm that, for sufficiently weak confinement, i.e.,
$\eta_{\ast} \le 0.01$,  the wind is essentially unaffected by the 
magnetic field.
But for models within the range $1/10 < \eta_{\ast} < 10$, the field
has a significant influence on the wind.
For our main parameter study, the variations in $\eta_{\ast}$ are implemented 
solely through variations in the assumed magnetic field strength, 
with the stellar and wind parameters fixed at 
values appropriate to a typical OB supergiant, e.g.  $\zeta$ Pup,
as given in Table 1.
Following this, we briefly (\S 4.4) compare flow configurations with 
identical confinement parameter, $\eta_{\ast}$, but obtained with different 
stellar and wind parameters.

\subsection{Time Relaxation of Wind to a Dipole Field}

\begin{figure}
\begin{center}
\plotone{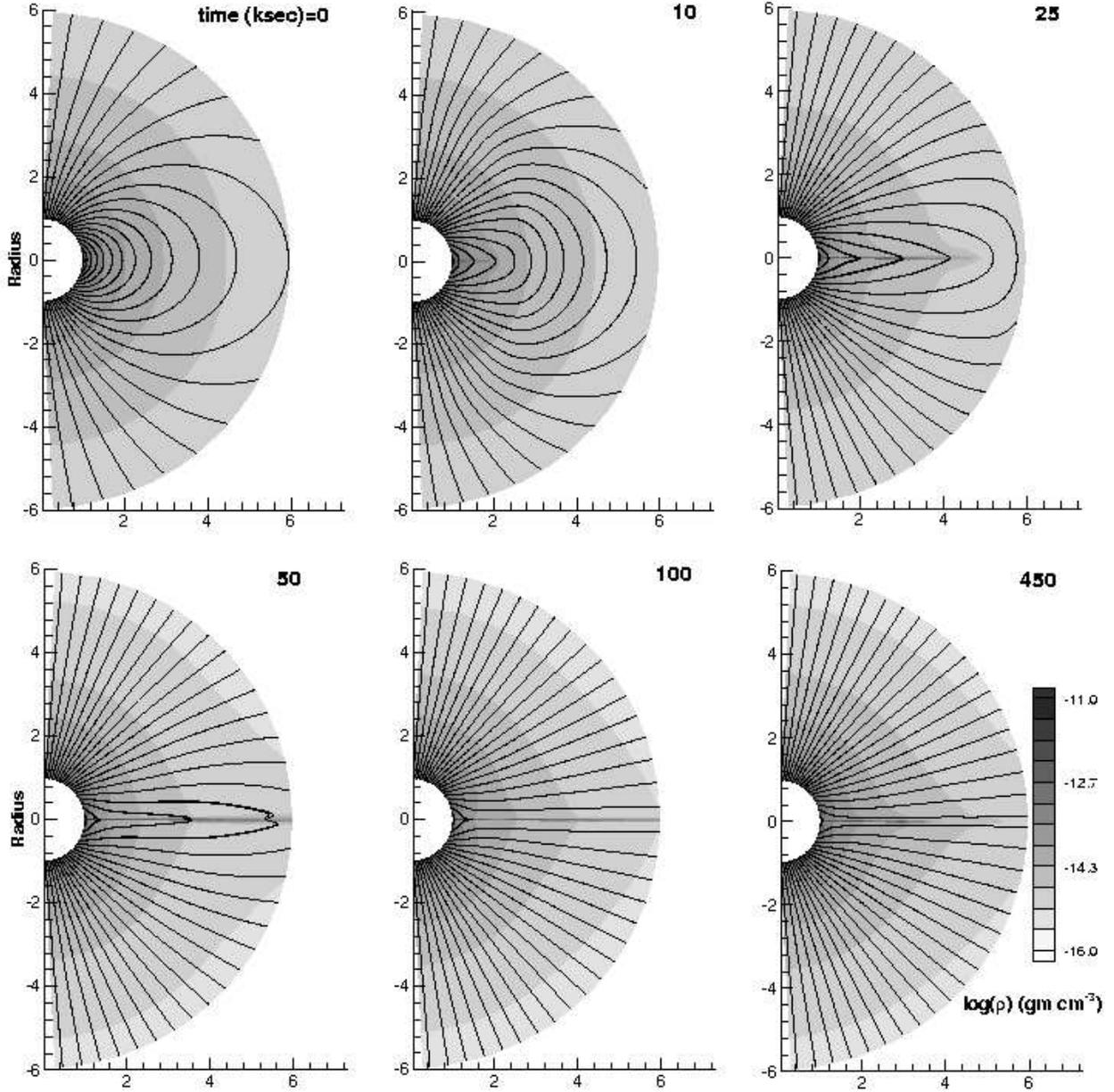}
\end{center}
\caption{
Snapshots of density (as logarithmic grayscale) and magnetic field (lines) 
at the labeled time intervals starting from the initial condition of a 
dipole field superposed upon a spherically symmetric outflow, for
a case of mderate magnetic confinement $\eta_{\ast}=\sqrt(10)$
($B_{Pole}=520G$).
The intervals of field lines emanating from the star are chosen to
preserve the relationship of field-line density with the strength of 
magnetic field.
}
\label{fig2}
\end{figure}

As noted above, we study the dynamical competition between field and 
wind by evolving our MHD simulations from an initial condition at time $t=0$, 
when a dipole  magnetic field is suddenly introduced into a previously 
relaxed, 1D spherically symmetric CAK wind.
For the case of moderately strong magnetic confinement, $\eta_{\ast} = 
\sqrt{10}$ ($B_o = 520$~G), figure \ref{fig2} illustrates the evolution of
magnetic field (lines) and density (gray scale) at fixed time snapshots,
$t=$0, 10, 25, 50, 100, and 450~ksec.
Note that the wind outflow quickly stretches the initial dipole field
outward, opening up closed magnetic loops and eventually forcing the
field in the outer wind into a nearly radial orientation.
This outward field-line stretching implies a general
enhancement of the magnetic field magnitude in the outer wind, as
evidenced in figure \ref{fig2} by the increased density of field
lines at the later times.
This global relaxation of field and flow is completed within the
first $50-100$~ksec, corresponding to about 2-4 times the characteristic flow
crossing time, $t_{flow} = 5 \Rstar/\vinf \simeq 25$~ksec.

To ascertain the asymptotic behavior of flows with various magnetic 
confinement parameters $\eta_{\ast}$, we typically run 
our simulations to a time, $t=450$~ksec, that is much longer (by 
factor $\sim$18) than this characteristic flow time.
Generally we find that, after the initial relaxation period of 
$\sim 50-100$~ksec,
the outer wind remains in a nearly stationary configuration, with 
nearly steady, smooth outflow along open field lines. 
However, for cases with sufficiently strong magnetic field
$\eta_{\ast} > 1$, confinement by closed loops near the equatorial 
surface can lead to quite complex flows, with persistent, intrinisic variability.
Before describing further this complex structure near the stellar 
surface,  let us first examine the simpler, nearly steady flow 
configurations that result in the outer wind.

\subsection{Global Wind Structure for Strong, Moderate, and Weak  Fields}

\begin{figure}
\epsscale{0.8}
\begin{center}
\plotone{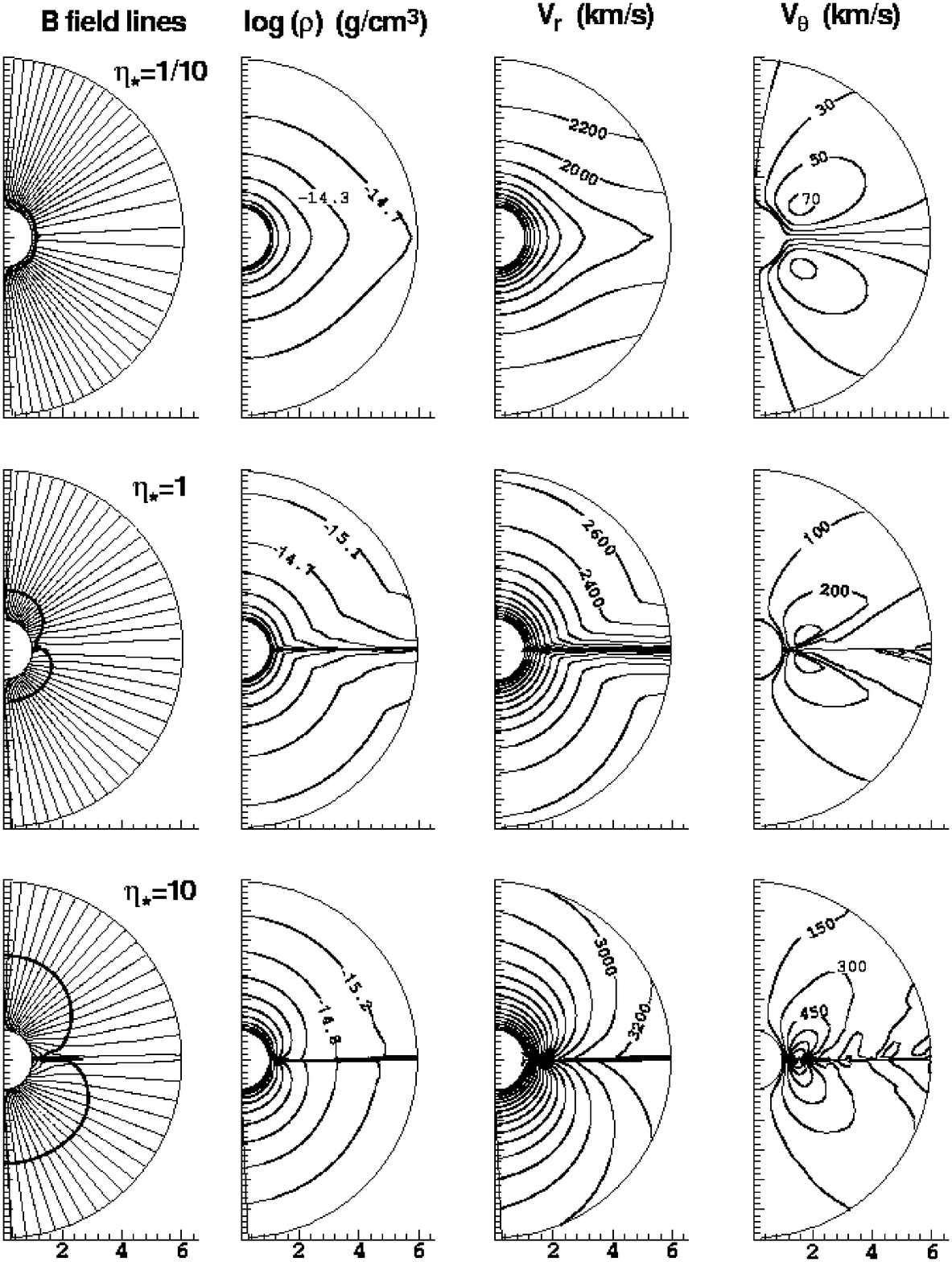}
\end{center}
\caption
{
Comparison of overall properties at the final simulation 
time ($t=450$~sec) for 3 MHD models, chosen to span a range of 
magnetic confinement from 
small (top row; $\eta_{\ast}=1/10$), 
to medium (middle row; $\eta_{\ast}=1$),
to large (bottom row; $\eta_{\ast}=10$).
The leftmost panels show magnetic field lines, together with the 
location (bold contour) of the Alfven radius, where the radial flow speed 
equals the Alfven speed.
From left to write, the remaining columns show 
contours of log(density), radial velocity, and latitudinal velocity.
}
\label{fig3}
\end{figure}

Figure \ref{fig3} illustrates the global configurations of magnetic field, 
density, and radial and latitudinal components of velocity at the 
final time snapshot, $t=450$~ksec after initial introduction 
of the dipole magnetic field. 
The top, middle, and bottom rows show respectively results for a weak,
moderate, and strong field, characterized by confinement parameters of
$\eta_{\ast} =1/10$, $1$, and $10$.

For the weak magnetic case $\eta_{\ast}=1/10$, note that the flow 
effectively extends the field to almost a purely radial configuration everywhere.
Nonetheless, even in this case the field still has a noticeable influence, 
deflecting the flow slightly toward the magnetic equator 
(with peak latitudinal speed $\max(v_\theta) \simeq 70$~km/s)  
and thereby leading to an increased  
density and a decreased radial flow speed in the equatorial region.

For the moderate magnetic strength $\eta_{\ast}=1$, this equatorward 
deflection becomes more pronounced, with a faster 
latitudinal velocity component ($\max(v_\theta ) \simeq 300$~km/s),
and a correspondingly stronger equatorial change in density and radial 
flow speed.
The field geometry shows moreover a substantial nonradial tilt near 
the equatorial surface.

For the strong magnetic case  $\eta_{\ast}=10$, the near-surface fields 
now have a closed-loop configuration out to a substantial fraction of 
a  stellar radius above the surface.  
Outside and well above the  closed region, the flow is quasi-steady, 
though now with substantial channeling
of material from higher latitudes toward the magnetic equator,
with $ \max(v_\theta) > 500$~km/s, even outside the closed loop.
This leads to a very strong flow compression,  and thus to a
quite narrow equatorial ``disk'' of dense, slow outflow.

This flow configuration is somewhat analagous to the ``Wind Compressed
Disk'' model developed for non-magnetic, rotating winds (Bjorkman and
Cassinelli 1993; Owocki, Cranmer, and Blondin 1994).
Indeed, it was
already anticipated as a likely outcome of magnetic channeling, e.g.
by BM97b, and in the ``WC-Fields'' paradigm
described by Ignace, Bjorkman and Cassinell (1998).
It is also quite analogous to what is found for strong field 
channeling in other types of stellar wind (Matt et al. 2000),
including even the solar wind (Keppens and Goedbloed 1999 2000).

\subsection{Variability of Near-Surface Equatorial Flow}

\begin{figure}
\epsscale{1.0}
\begin{center}
\plotone{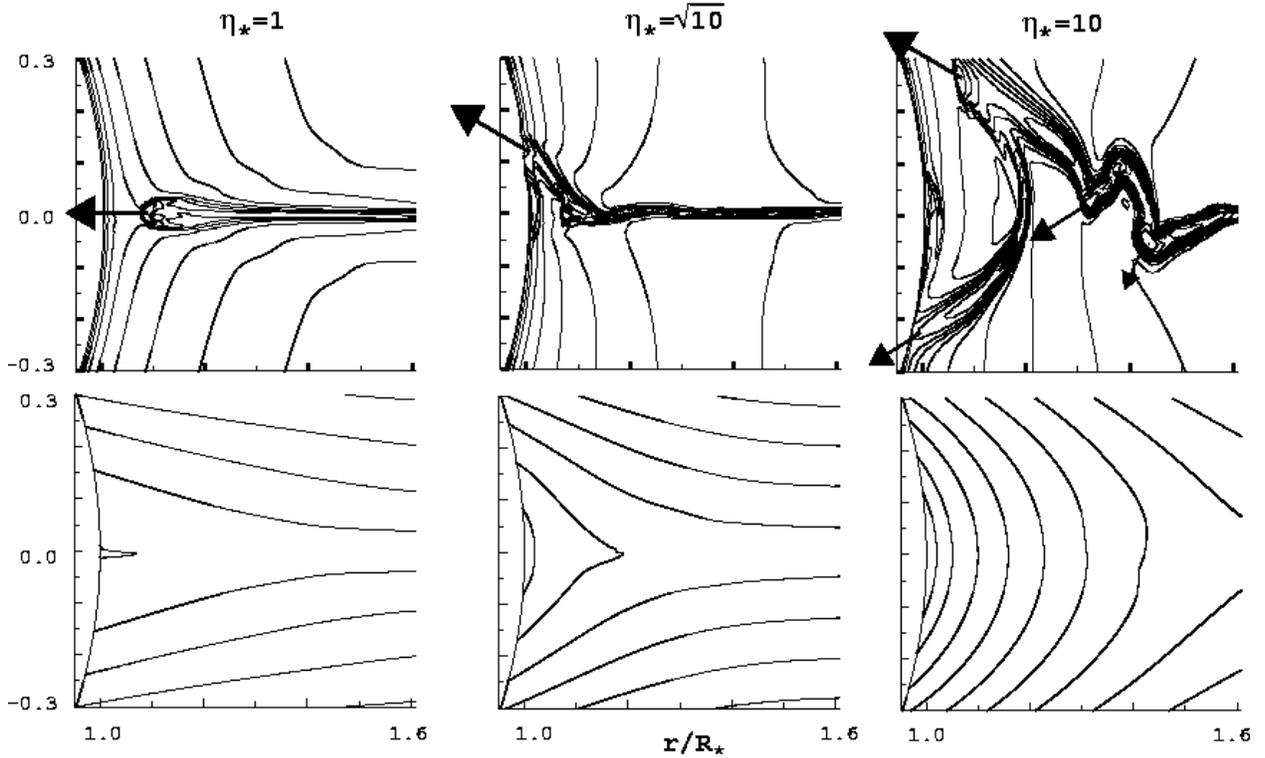}
\end{center}
\caption
{
Contours of log(density) (upper row) and magnetic field lines (lower 
row) for the inner, magnetic-equator regions of 
MHD models with moderate ($\eta_{\ast}=1$; left),
strong ($\eta_{\ast}=\sqrt{10}$; middle), and
strongest ($\eta_{\ast}=10$; left) magnetic confinement, 
shown at a fixed,  arbitary time snapshot well after ($t \ge 400$~ksec) 
the initial condition.
The arrows represent the direction and magnitude of the mass flux, and
show clearly that the densest structures are undergoing an infall back
onto the stellar surface.
For the moderate magnetic confinement $\eta_{\ast}=1$, this infall is directly 
along the equator, but for the higher confinements $\eta_{\ast}=\sqrt{10}$
and $10$, the equatorial compressions that form at larger radii are 
deflected randomly toward the north or south as they fall in toward
the closed field near the surface.
The intent here is to illustrate how increasing magnetic confinement 
leads to an increasing complexity of flow and density structure within 
closed magnetic loops.
This complexity is most vividly illustrated in the time animations available in
the electronic version of the paper.
}
\label{fig4}
\end{figure}

In contrast to this relatively steady, smooth nature of the outer wind, 
the flow near the star can be quite structured and variable in the 
equatorial regions.
For the strong magnetic field case $\eta_{\ast} = 10$, 
the complex structure of the flow within the closed magnetic
loops near the equatorial surface is already apparent even in the
global contour plots in the bottom row of figure \ref{fig3}.
To provide a clearer illustration of this variable flow structure,
Figure \ref{fig4} zooms in on the near-star equatorial region,  comparing
density (upper row, contours), mass flux (arrows), and field lines 
(lower row) at an arbitray time snapshot long after the initial condition
at an arbitray time snapshot long after the initial condition
($t >400$~ksec), for three models with magnetic 
confinement numbers $\eta_{\ast}=$ 1, $\sqrt{10}$, and 10.

For the case of moderate magnetic field with $\eta_{\ast}=1$, note the 
appearance of the high-density knot at a height $\sim 0.3 \Rstar$ 
above the equatorial surface.
As is suggested from the bow-shaped front on the inward-facing side of
this knot, it is flowing {\it inward}, the result of a general infall
of material that had been magnetically channeled into an equatorial
compression, and thereby became too dense for the radiative
line-driving to maintain a net outward acceleration against the inward
pull of the stellar gravity.
This again is quite analogous to the inner disk infall found in dynamical 
simulations of rotationally induced Wind Compressed Disks (Owocki, 
Cranmer, and Blondin 1994).

Animations of the time evolution for this case show that such dense
knots form repeatedly at semi-regular intervals of about 200~ksec. 
A typical cycle begins with a general building of the equatorial density
through the magnetic channeling and equatorial compression of wind
outflow from higher latitudes.  
As the radiative driving weakens with the increasing density, the
equatorial outflow first decelerates and then reverses into an inflow
that collects into the dense knot.  
Once the equatorial density is emptied by the knot falling onto the
surface, the build-up begins anew, initiating a new cycle.

For the case of somewhat stronger field with $\eta_{\ast}=\sqrt{10}$, the
snapshot at the same fixed time $t=400$~ksec again shows evidence for 
an infalling knot, except that now this has been forced by the 
magnetic tension of an underlying,  closed, equatorial loop to slide to one 
hemisphere (in this case north), instead of falling directly upon the
equatorial stellar surface.
Animations of this case show a somewhat more irregular repetition, 
with knots again forming about every 200~ksec, but randomly sliding
down one or the other of the footpoint legs of the closed  equatorial loop.
It is interesting that, even though our simulations are formally 
symmetric in the imposed conditions for the two hemispheres, this 
{\it symmetry is spontaneously broken} when material from the overlying 
dense equatorial flow falls onto, and eventually off of, the top of the 
closed magnetic loop.

The strongest field case with $\eta_{\ast}=10$ shows a much more extensive
magnetic confinement, and accordingly a much more elaborate configuration
for material re-accretion onto the surface. 
Instead of forming a single knot, the equatorially compressed material 
now falls back as a complex ``snake'' of dense structure that breaks up 
into a series of dense knots, draining down the magnetic loops toward both
northern and southern footpoints.
In the time animations for as long as we have run this case, there is 
no clear repetition time, as the formation and infall of knots and
snake-segments are quite random, perhaps even chaotic.

It is worth noting here that in the two stronger magnetic cases, the 
closed loops include a region near the surface for which the field 
is so nearly horizontal that it apparently 
{\it  inhibits} any net upflow.
As a consequence, these loops tend to become quite low density.
This has an unfortunate practical consequence for our numerical 
simulations, since the associated increase in the Alfven speed 
requires a much smaller numerical time step from the standard Courant 
condition.
Thus far this has effectively limited our ability to run such strong field 
models for the very extended time interval needed for clearer
definition of the statistical properties of the re-accretion process.

In contrast to this effective inhibition of radial outflow by the
nearly horizontal field in the central regions of a closed loop,
note that in the outer portion of the loop, where the field is 
more vertical, the radial line-driving is able to initiate a supersonic
flow up along the loop.
But when this occurs along a field line that is still closed, the 
inevitable result is that material from opposite footpoints is 
forced to collide near the top of the loop.
This effectively halts the outflow for that field line, with the
accumulating material near the top of the loop supported by both
the magnetic tension from below and the ram pressure of incoming
wind from each side.

As the density builds, maintaining this support against gravity 
becomes increasingly difficult.
For the moderate field strength, the material from all such closed
flow tubes accumulates into a knot whose weight forces the loop top
to  buckle inward, first nearly symmetrically but eventually off toward one
side, allowing the material to re-accrete toward that footpoint at
the surface.
For the strongest field, the loops tend to remain nearly rigid, keeping
the material from distinct closed flow tubes separate and suspended
at loop tops with a range of heights, until this line of 
material finally breaks up into segments (the ``snake'') that 
fall to either side of the rigid re-accretion tubes.

It is interesting to contrast this inferred outflow and re-accretion 
in the magnetic equator of a line-driven wind with what 
occurs in the solar coronal expansion.
For the solar wind, acceleration to supersonic speeds typically occurs
at a height of several solar radii above the surface.
As such, magnetic loops that typically close within such heights
can generally maintain the gas in a completely {\it hydrostatic}
stratification.
By contrast, in line-driven winds supersonic speeds are typically 
achieved very near the stellar surface, within about $0.1 \Rstar$.
A closed loop that starts from a radially oriented footpoint thus 
simply guides this line-driven outflow along the loop,
instead of confining a hydrostatic stratification. 
For a strong field with sufficiently high loop top, the eventual shock 
collision can have velocity jumps that are a substantial fraction 
of the  wind terminal speed, e.g. $\sim 1000$~km/s.
In the present isothermal simulations, the heating from such shocks 
is assumed to be radiated away over a narrow, unresolved cooling layer.
In the discussion in \S 5.3.4, we estimate some general properties of the
associated X-ray emission.

\begin{table}
\renewcommand{\arraystretch}{1.1}
\begin{center}
\caption{Summary Table}
\begin{tabular}{lrrrrrrrr}
\hline
\hline
&\multicolumn{7}{c}{Model}\\
\cline{2-9}
 & & & & & & &\multicolumn{2}{c}{$\eta_{\ast}=\sqrt{10}$} \\
\cline{8-9}
Quantity&$\eta_{\ast}=0$&$\eta_{\ast}=\frac{1}{10}$&$\eta_{\ast}=\frac{1}{\sqrt{10}}$&
$\eta_{\ast}=1$&$\eta_{\ast}=\sqrt{10}$&$\eta_{\ast}=10$&low $\dot{M}$&$\theta^1$ Ori C\\
\hline

$\alpha$&0.6&0.6&0.6&0.6&0.6&0.6&0.6&0.5\\
$\bar{Q}$&500&500&500&500&500&500&20&700\\
$\delta$&0.0&0.0&0.0&0.0&0.0&0.0&0.0&0.1\\
$R_{\ast}(10^{12}cm)$&1.3&1.3&1.3&1.3&1.3&1.3&1.3&0.5\\
$B_{Pole}$(G)&0&93&165&295&520&930&165&480\\
$\rho_0$($10^{-11} gm \,cm^{-3}$)&4.3&4.3&4.3&4.3&4.3&4.3&0.54&2.8\\
max($v_r)(km\; s^{-1})$&2300&2350&2470&2690&2830&3650&2950&2620\\
max($v_{\theta})(km\; s^{-1})$&0&70&150&300&400&1200&400&450\\
$\dot{M}_{net}$($10^{-6}M_{\odot}~yr^{-1}$)&2.6&3.0&2.8&2.5&2.2&1.8&0.22&0.3\\
\hline

\end{tabular}
\end{center}
\end{table}

\subsection{Comparing Models with Different Stellar Parameters but 
Fixed $\eta_{\ast}$}

\begin{figure}
\epsscale{1.0}
\begin{center}
\plotone{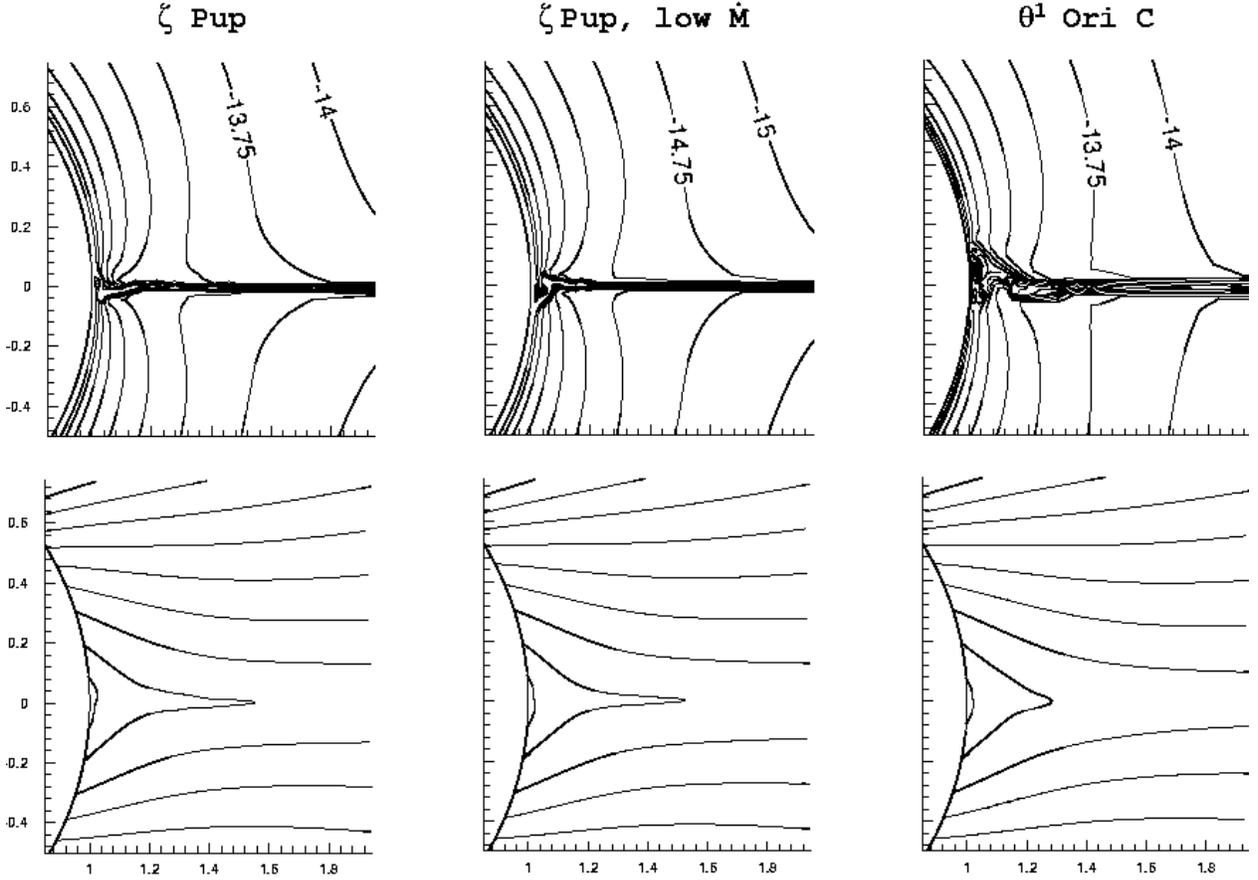}
\end{center}
\caption
{
Contours for log of density (upper row) and magnetic fields (lower 
row) in the inner regions of 3 MHD models with a fixed magnetic 
confinement parameter $\eta_{\ast}=\sqrt{10}$, but obtained using 
different stellar and wind parameters, chosen to correspond to an O-type
supergiant like $\zeta$~Pup with normal (left) or factor-ten lower 
mass loss rate (middle), or to a late-O giant like $\theta^1$~Ori~C
(right).
The overall similarity of the three models illustrates the degree to 
which the global configuration of field and flow depends mainly on just
the combination of stellar, wind, and magnetic properties that define
[via eqn. (\ref{wmcpdef}] the magnetic confinement parameter $\eta_{\ast}$.
}
\label{fig5}
\end{figure}

The models above use different magnetic field
strengths to vary the magnetic confinement
for a specific O supergiant star with fixed stellar and wind parameters.
To complement that approach, let us briefly examine models
with a fixed confinement parameter
$\eta_{\ast}=\sqrt{10}$, but now manifest through different
stellar and/or wind parameters.
Specifically, figure \ref{fig5} compares density contours (top) and 
field lines (bottom) for our standard $\zeta Pup$
model (center panels), with a model in which the mass loss rate 
is reduced by a  factor 10 (left panels), and also with a model based 
on entirely different stellar and wind parameters, intended roughly to 
represent the O7 star $\theta^1$~Ori~C (right panels).
(Detailed parameter values are given in Table 1.)

Note that all three models have very similar overall {\it form} 
in both their density contours and field lines, even though the 
associated magnitudes vary substantially from case to case.
This similarity of structure for models with markedly different 
individual parameters, but configured to give
roughly equal $\eta_{\ast}$,  thus further reinforces the notion that 
this confinement parameter is the key determinant in fixing the
overall competition between field and flow.

\section{Analysis and Discussion}

\subsection{Comparison of MHD Simulations with Heuristic Scaling
Estimates}

The above results lend strong support to the general idea,
outlined in \S 3.1, that the overall effect of a magnetic field in
channeling and confining the wind outflow depends largely on the single
magnetic confinement parameter $\eta_{\ast}$.
Let us now consider how well these MHD simulation results correspond
to the heuristic estimates for the Alfven radius $R_A$ and
magnetic closure colatitude $\theta_A$ defined in \S 3.2.

\begin{figure}
\epsscale{0.75}
\begin{center}
\plotone{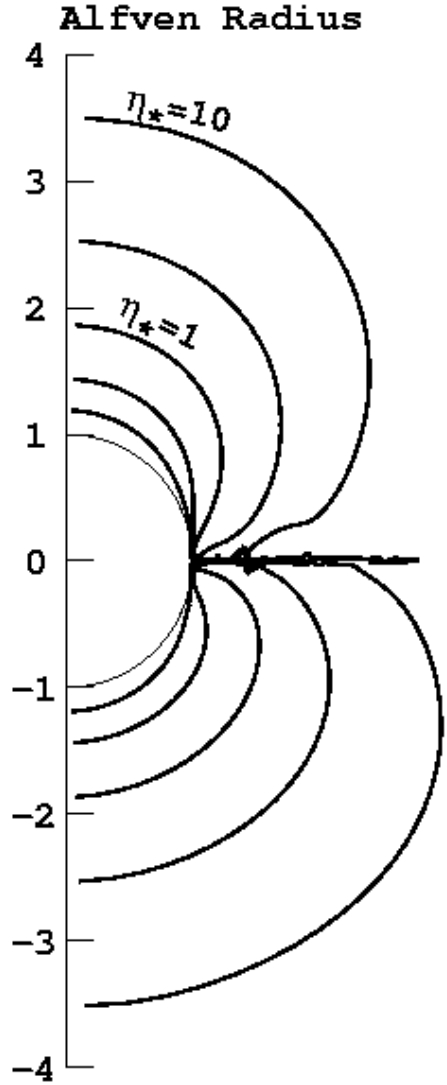}
\end{center}
\caption{
Contours of the Alfven radius in MHD models with confinement
parameters (from inside-out) $\eta_{\ast} =$~1/10, $1/\sqrt{10}$, 
1,  $\sqrt{10}$, and $10$.
}
\label{fig6}
\end{figure}

Figure \ref{fig6} plots contours of the Alfven radius obtained in the numerical
MHD simulations with various $\eta_{\ast}$.
Reflecting the stronger field and so higher Alfven speed, 
the models with larger confinement parameter have a higher Alfven 
radius.
Note, moreover, that for all cases the Alfven radius generally decreases
toward the equator.
In part, this just reflects the Alfven speed associated with
the dipole surface magnetic field, which has a lower 
strength near the magnetic equator.

But the comparison in figure \ref{fig6} shows a systematic discrepancy between
the curves showing the expected Alfven radius from this dipole model and 
the points showing the actual MHD results.
Specifically, the dipole model underestimates the MHD Alfven radius 
over the pole, and overestimates it at the equator.

\begin{figure}
\begin{center}
\plotone{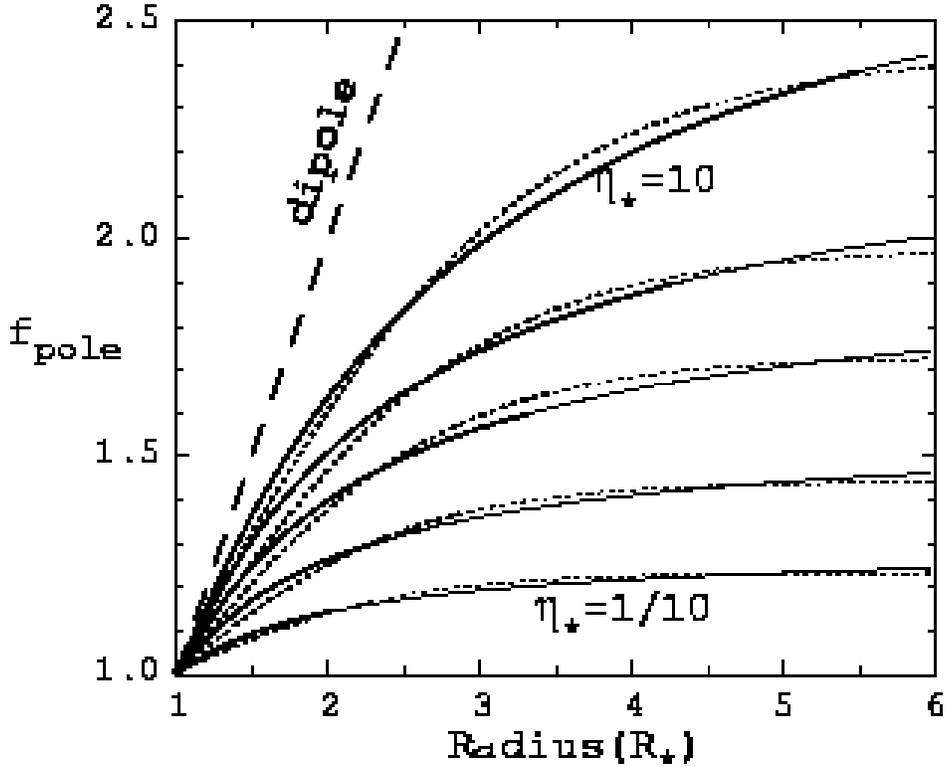}
\end{center}
\caption
{ 
The measure of faster-than-$r^2$ decline of the polar magnetic field, 
as represented by the function $f_{pole}$ defined in eqn. (\ref{fpoledef}), 
and applied to the 5 MHD simulations with magnetic confinement ranging 
from strong  ($\eta_{\ast}=10$; uppermost curve) 
to weak ($\eta_{\ast}=10$; lowermost curve).
For the ideal MHD cases here of field-frozen flow, this also 
represents the degree of faster-than-$r^2$ expansion of flow tube area.
The dotted curves plot the heuristic area-expansion
function defined by Kopp and Holzer [eqn. (\ref{fkhdef}) here], 
with $R_1=1~R_\ast$, and the parameters 
$f_{max}=$~2.43, 1.98, 1.73, 1.44 and 1.23, 
and 
$\sigma/R_\ast=$~1.13, 0.98, 0.89, 0.79, and 0.73,
chosen to best fit to the five cases from $\eta_{\ast}=$10 to 1/10.
The dashed curve shows the variation for a pure dipole field.
}
\label{fig7}
\end{figure}

For the polar wind, this can be understood as a consequence of the
radial stretching of the field.
Figure \ref{fig7} plots the radial variation of the polar field ratio 
\begin{equation}
f_{pole} (r) \equiv 
{ R_\ast^2 
B_o
\over
r^2 
B(r,0) 
}
\label{fpoledef}
\end{equation}
for the various magnetic confinement parameters $\eta_{\ast}$.
For comparison, a dipole field (with $B \sim r^{-q}$ and $q=3$) 
would just give a straight line of unit slope (dashed line),
whereas a pure monopole, radial field (with $q=2$) would give 
a horizontal line at value unity, $f_{pole}=1$.

%
    
The results show that the MHD cases are intermediate between these
two limits. 
For the weakest confinement $\eta_{\ast} = 1/10$, the curve bends toward 
the horizontal at quite small heights, reflecting how even the inner
wind is strong enough to extend the polar field into a nearly 
radial orientation and divergence.
For the strongest confinement $\eta_{\ast} = 10$, the field divergence
initially nearly follows the dashed line for a dipole ($q=3$), but then
eventually also bends over as the wind ram pressure overwhelms the 
magnetic confinement and again stretches the field into a nearly radial
divergence.
The intermediate cases show appropriately intermediate trends, but in 
all cases it is significant that the radial decline in field strength 
is generally less steep than for a pure dipole, i.e. $q<3$.
The dotted line in figure \ref{fig1} indeed shows that the MHD results for the
polar Alfven radii of the various confinement cases are in much better
agreement with a simple scaling that assumes a radial decline 
(power index $q=2.6$) 
that is intermediate between the dipole ($q=3$) and monopole (i.e. radial
divergence, $q=2$) limits.

Overall then, at the poles the radial stretching of field by the outflowing
wind has the net effect of reducing the radial decline of field, and thus 
increasing the Alfven radius over the value expected from the simple 
dipole estimate of eqn. (\ref{radef}).

By contrast, at the equator this radial stretching has a somewhat
opposite effect, tending to {\it remove} the predominantly {\it horizontal} 
components of the equatorial dipole field, and thus leading to
a lower equatorial field strength and so also a lower assoicated
Alfven radius, relative to the simple dipole analysis of \S 4.2.
For example, for the lowest confinement case $\eta_{\ast}=1/10$, the
field is extended into a nearly radial configuration almost right from
the stellar surface, as shown by the top left panel of figure 
\ref{fig3};
the equatorial polarity switch of this radial field thus implies a vanishing 
equatorial Alfven speed, which thus means that contours of Alfven
radius must bend sharply inward toward the surface near the equator.
For the strongest confinement case $\eta_{\ast}=10$, the near-surface horizontal 
field within closed magnetic loops about the equator remains strong enough 
to resist this radial stretching by wind outflow;
but the faster radial fall-off in magnetic vs. flow energy means that the 
field above  these closed loops is eventually stretched outward into a radial 
configuration, thus again leading to a vanishing equatorial field and 
an associated inward dip in the Alfven radius.


This overall dynamical lowering of the equatorial strength of magnetic field 
further means that the latitudinal extents of
closed loops in full MHD models are generally below what is predicted
by the simple dipole estimate of eqn. (\ref{tcdef}).
Thus, in previous semi-analytic models of BM97b, which effectively 
assume this type of dipole scaling, a somewhat larger surface field is
needed to give the assumed overall extent of magnetic confinement.

\subsection{Effect of Magnetic Field on Mass Flux and Flow Speed}

\begin{figure}
\epsscale{1.0}
\begin{center}
\plotone{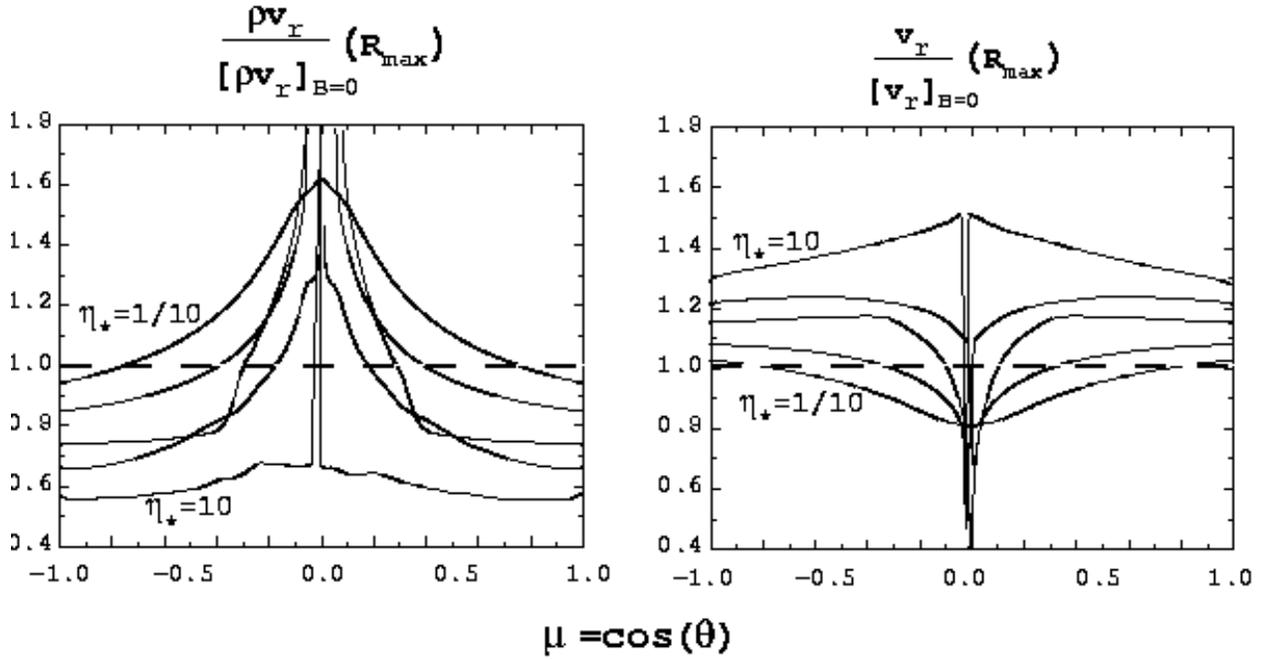}
\end{center}
\caption
{
The radial mass flux density $\rho v_r$ (left) and
radial flow speed $v_r$ (right) at the maximum model radius $R_{max}=6 R_\ast$,
normalized by values in the corresponding non-magnetic model,
and plotted versus the cosine of the colatitude, $\cos(\theta)$, for
the final time snaphsot ($t=450$~ksec) in the
5 MHD models with magnetic confinement parmeters 
$\eta_{\ast}=~1/10$, $1/\sqrt{10}$, 1, $\sqrt{10}$, and 10.
The horizontal dashed lines mark the unit values for the non-magnetic
($B=\eta_{\ast}=0$) case.
}
\label{fig8}
\end{figure}

Two key general properties of spherical, non-magnetic stellar winds are the
mass loss rate ${\dot M} \equiv 4 \pi \rho v_r r^2$ and 
terminal flow speed $v_\infty$.
To illustrate how a stellar magnetic field can alter these properties
for a line-driven wind, figure \ref{fig8} shows the outer boundary 
($r=R_{max}$) values of the 
radial velocity, $v_r(R_{max},\theta)$, 
and 
radial mass flux density, $\rho(R_{max},\theta) v_r(R_{max},\theta)$,
normalized by the values (given in Table 1) for the non-magnetic, 
spherically symmetric wind case, and plotted as a function of
$\mu=\cos(\theta)$ for each of our simulation models with various 
confinement parameters $\eta_{\ast}$.

There are several noteworthy features of these plots.
Focussing first on the mass flux, 
note that in all models the tendency of the field to divert 
flow toward the magnetic equator leads to a general increase in mass flux
there, with this equatorial compression becoming narrower with increasing field
strength, until, for the strongest field, it forms the spike
associated with an equatorial disk.
This higher equatorial mass flux is associated with a higher  
density, since the equatorial flow speeds are always lower, quite
markedly so for the dense, slowly outflowing disk of the strong
field case.

Table 1 lists the overall mass loss rates, obtained by integration of
these curves over the full range $-1<\mu<1$. 
For the strong field case, the mass loss is reduced relative to the 
non-magnetic ${\dot M}$, generally because the magnetic confinement and tilt
of the inner wind outflow has effectively inhibited some of the base mass flux.
Curiously, for the weakest magnetic confinement case $\eta_{\ast}=1/10$,
there is actually a modest overall {\it increase} in the mass loss.
The reasons for this are not apparent, and will require further 
investigation.

\subsubsection{Role of Rapid Areal Divergence in Enhancing Wind Flow 
Speed}

The right panel in figure \ref{fig8} shows that the wind flow speed
over the poles is enhanced relative to a spherical wind.
This is quite reminiscent of the high-speed polar flow
in the solar wind (Smith, Balogh, Forsyth, \& McComas 2001; Horbury \& Balogh 2001),
 which is generally understood to emanate 
from polar coronal holes of open magnetic field 
(Zirker 1977).
Modeling of such high-speed solar wind has emphasized the important role of
faster-than-radial area divergence in such open field regions
(Kopp and Holzer 1976; Holzer 1977; Wang and Sheeley 1990).

Indeed, based on this solar analogy, MacGregor (1988) analyzed the 
effect of such rapid divergence on a line-driven stellar wind, assuming
a simple 1D, radially oriented flow tube, as expected near the 
polar axis of an open magnetic field.
He concluded that, because the line-driving acceleration scales inversely
with density [$g_{lines} \sim 1/\rho^{\alpha}$; see eqn. 
(\ref{glines})], the lower density associated with faster divergence
would lead to substantially faster terminal speeds, up to a 
{\it factor three faster} than in a spherical wind, for quite
reasonable values of the assumed flow divergence parameters.
By comparison, the polar flow speed increases found in our full MHD models
here are much  more modest, about 30\% in even the strongest field case, 
$\eta_{\ast}=10$.

Because field and flow lines are locked together in the ideal MHD cases
assumed here, the quantity $f_{pole}$, defined in eqn. (\ref{fpoledef}) 
and plotted in figure \ref{fig7},
actually also represents just this non-radial flow divergence factor for the 
polar flow. 
It is thus worthwhile to compare these dynamically computed divergence 
factors to the divergence assumed by MacGregor 
(1988), which was based on a heuristic form introduced originally 
by Kopp and Holzer (1976) for the solar case
\begin{equation}
    f(r) = 
    { f_{max} \exp[(r-R_1)/\sigma]+ 1 - (f_{max}-1) \exp[(R_\ast-R_1)/\sigma]
    \over
    \exp[(r-R_1)/\sigma] + 1
    } \, .
\label{fkhdef}
\end{equation}
Specifically, these previous analyses generally assumed that the rapid
divergence would be confined to a quite narrow range of radius 
($\sigma = 0.1 R_\ast$) centered on some radius
($R_1 = 1.25-2.5 \, R_\ast$)
distinctly above the stellar surface radius $R_\ast$.
By comparison, our dynamical simulations indicate the divergence is 
generally most-rapid right at the wind base (implying $R_1 \approx R_\ast$), 
and extends over a quite large radial range (i.e., $\sigma > 1 R_\ast$).
On the other hand, the MacGregor (1988) assumed values of the asymptoptic net divergence, 
$f_{max}= 1.25-2.0$, are quite comparable the divergence factors found
at the outer boundary of our MHD simulation models, 
$f_{pole} (R_{max}) = 1.25-2.5$.

These detailed differences in radial divergence
do have some effect on the overall wind acceleration, and thus on the 
asymptotic flow speed.
But it appears that the key reason behind the MacGregor (1988) prediction of 
a very strong speed enhancement was the neglect there of the finite-disk
correction factor for the line-force 
(Friend and Abbott 1986; Pauldrach, Puls, and Kudritzki 1986).
With this factor included, and using the MHD simulations to define 
both the divergence and radial tilt-angle of the field and flow, we 
find that a simple flow-tube analysis is able to explain quite well our 
numerical simulation results for not only the polar speed, but
also for the latitudinal scaling of both the speed and mass flux.
In particular, we find that the even stronger increase in flow speed 
seen at mid-latudes ($ 1/4 < |\mu| < 3/4$) in the strongest field
model ($\eta_{\ast}=10$) does not reflect any stronger divergence factor,
but rather is largely a consequence of a {\it reduced} base mass flux
associated with a nonradial tilt of the source flow near the stellar
surface.
As the flow becomes nearly radial somewhat above the wind base,
the lower density associated with the lower mass flux implies a stronger
line-acceleration and thus a faster terminal speed along these 
mid-latitude flow tubes.
Further details of these findings will be given in future paper.

\subsection{Observational Implications of these MHD Simulations}

\subsubsection{UV Line-Profile Variability}

It is worth emphasizing here that these dynamical results for the
radial flow speed have potentially important implications for 
interpreting the observational evidence for wind structure and
variability commonly seen in UV line profiles of hot stars 
(e.g., the so-called Discrete Absorption Components; Henrichs et al. 1994; 
Prinja and Howarth 1986; Howarth and Prinja 1989).
In particular, an increasingly favored paradigm is that the inferred
wind structure may arise from Corotating Interaction Regions (CIRs)
between fast and slow speed wind streams. 
This requires a base perturbation mechanism to induce latitudinal
variations in wind outflow properties from the underlying, rotating
star (Mullan 1984; Cranmer and Owocki 1996).
Based largely on the analogy with solar wind CIRs -- for which the 
azimuthal  variations in speed are clearly associated with magnetic 
structure of the solar corona 
(Zirker 1977; Pizzo 1978) --, 
there has been a 
longstanding speculation that surface magnetic fields on hot stars 
could similarly provide the base perturbations for CIRs in line-driven 
stellar winds
(Mullan 1984; Shore \& Brown 1990, 
Donati 2001).

However, until now, one argument {\it against} this magnetic model for 
hot-star-wind structure was the expectation, based largely on the Macgregor
(1988) analysis, that a sufficiently strong field would likely lead to
anomalously high-speed streams, in excess of 5000~km/s, 
representing the predicted factor of two or more enhancement above the speed
for a non-magnetic wind
(Owocki 1994; BM97a).
%
%
%
%
%
By comparison, the wind flow speeds inferred quite directly from the 
blue edges of strong, saturated P-Cygni absorption troughs of UV lines
observed from hot stars show only a modest variation of a few hundred
km/s, with essentially {\it no evidence} for such extremely fast speeds
(Prinja et al. 1998).

The full MHD results here are much more in concert with this inferred speed 
variation, even for the strongest field model, for which the fastest 
streams are not much in excess of $\sim$3000~km/s. 
Moreover, in conjunction with the 
reduced flow speeds toward the magnetic equator, there is still quite sufficient
speed contrast to yield very strong CIRs, if applied in a rotating 
magnetic star with some substantial tilt between magnetic and rotation axes. 
Through extensions of the current 2D models to a full 3D configuration,
we plan in the future to carry out detailed simulations of winds from
rotating hot-stars with such a tilted dipole surface field, applying these 
specifically toward the interpretation of observed UV line profile variability.

\subsubsection{Infall within Confined Loops and Red-Shifted Spectral Features}

In addition to the slowly migrating discrete absorption components
commonly seen in the blue absorption troughs of P-Cygni profiles
of UV lines, there are also occasional occurences of {\it redward} 
features in either absorption (e.g.,  in $\tau$~Sco; Howk et al. 2000)
or emission (e.g., in $\lambda$~Eri and other Be or B supergiant stars; 
Peters 1986; Smith, Peters, and Grady 1991; Smith 2000; Kaufer 2000).
Within the usual context of circumstellar material that is
either in an orbiting disk or an outflowing wind, such redshifted spectral 
features have been difficult to understand, since they require material
flowing {\it away }from the observer, either in absorption 
against the stellar disk, or in emission from an excess of receding material 
radiating from a volume not occulted by the star.
In general this thus seems to require material {\it infall} back 
toward the star and onto the surface. Indeed, there have been 
several heuristic models that have postulated such infall might
result from a stagnation of the wind outflow, for example due
to clumping (Howk et al. 2000), or decoupling of the driving ions 
(Porter and Skouza 1999).

In this context, the dynamical MHD models here seem to provide 
another, quite natural explanation, namely that such infall is 
an inevitable outcome of the trapping of wind material within
close magnetic loops whenever there sufficiently strong wind magnetic 
confinement, $\eta_{\ast} \ge 1$.
In principal, such interpretations of observed red-shifts in terms 
of infall within closed magnetic loops could offer the possibility of a
new, indirect diagnostic of stellar magnetic properties.
For example, the observed redshift speed could be associated with a 
minimium required loop height to achieve such a speed by gravitational
infall.
In future studies, we thus intend to generate synthetic line absorption
and emission diagnostics for these MHD confinement models, and compare
these with the above cited cases exhibiting redshifted spectral 
features.

\subsubsection{Effect on Density-Squared Emission}

In addition to such effects on spectral line profiles from scattering,
absorption, or emission lines, the extensive compression of material
seen in these MHD models should also lead to an overall enhancement of
those types of emission, both in lines and continuum, for which the
volume emission rate scales with the square of the density.
Specific examples include line emission from both collisional 
excitation or recombination, or free-free continuum emission in the
infrared and radio. In principle, the former might even lead to a
net emission above the continuum in the hydrogen Balmer lines, and thus
to formal classification as a Be star, even without the usual 
association of an orbiting circumstellar disk.
Such a mechanism may in fact be the origin for the occasional occurence
of hydrogen line emission in slowly rotating B stars, most notably
$\beta$~Ceph, for which there has indeed now been a positive detection of
a tilted dipole field of polar magnetic around 300 G 
(Donati et al. 2001; Henrichs et al. 2000).
Again, further work will be needed to apply the dynamical MHD models here
toward interpretation of observations of density-square emissions from
hot stars that seem likely candidates for substantial wind magnetic
confinement.

\subsubsection{Implications for X-ray Emission}

Particularly noteworthly among the potential observational 
consequences of these MHD models are the clear implications for interpreting 
the detection of sometimes quite hard, and even cyclically variable, 
X-ray emission from some hot stars.
As noted in the introduction, there have already been quite extensive 
efforts to model such X-ray emission within the context of a fixed 
magnetic field that channels wind flow into strong shock collisions (BM97a,b).
In contrast, while the isothermal MHD models here do not yet include the detailed 
energy balance treatment necessary for quantitative modeling of such
shocked-gas X-ray emission, they do provide a much more complete and 
dynamically consistent picture of the field and flow configuration associated 
with such magnetic channeling and shock compression.

Indeed, as a prelude to future quantitative models with explicit 
computations of the energy balance and X-ray emission, let us briefly
apply here an approximate analysis of our model results that can yield rough 
estimates for the expected level of compressional heating and 
associated X-ray production.
The central idea is to assume that, within the context of the present
isothermal models, any compressive heating that occurs is quickly 
balanced by radiative losses within a narrow, unresolved cooling layer
(Castor 1987; Feldmeier et al. 1997; Cooper 1994).
For  shock-type compressions with a sufficiently strong velocity jump,
this radiative emission should include a substantial component in the
X-ray bandpass.

Applying this perspective, we first identify within our simulation models 
locations of locally strong compressions, i.e. where there are 
substantial zone to zone decreases in flow speed along the direction of
the flow itself. Taking into account that the quadratic viscosity 
within the Zeus code typically spreads any shocks over about 3 or 4 
zones, we can use this to estimate an associated total shock jump in the 
specific kinetic energy $\Delta v^2/2$.
We then apply the standard shock jump conditions
to obtain a correponding estimate of post-shock temperatures (BM97b),
\begin{equation}
    T_s \approx 2.7 \times 10^5 \, K ~ { - \Delta v^2 /2 \over (100 \, 
    km/s)^2 } \, .
\label{tsdef}
\end{equation}

\begin{figure}
\epsscale{0.65}
\begin{center}
\plotone{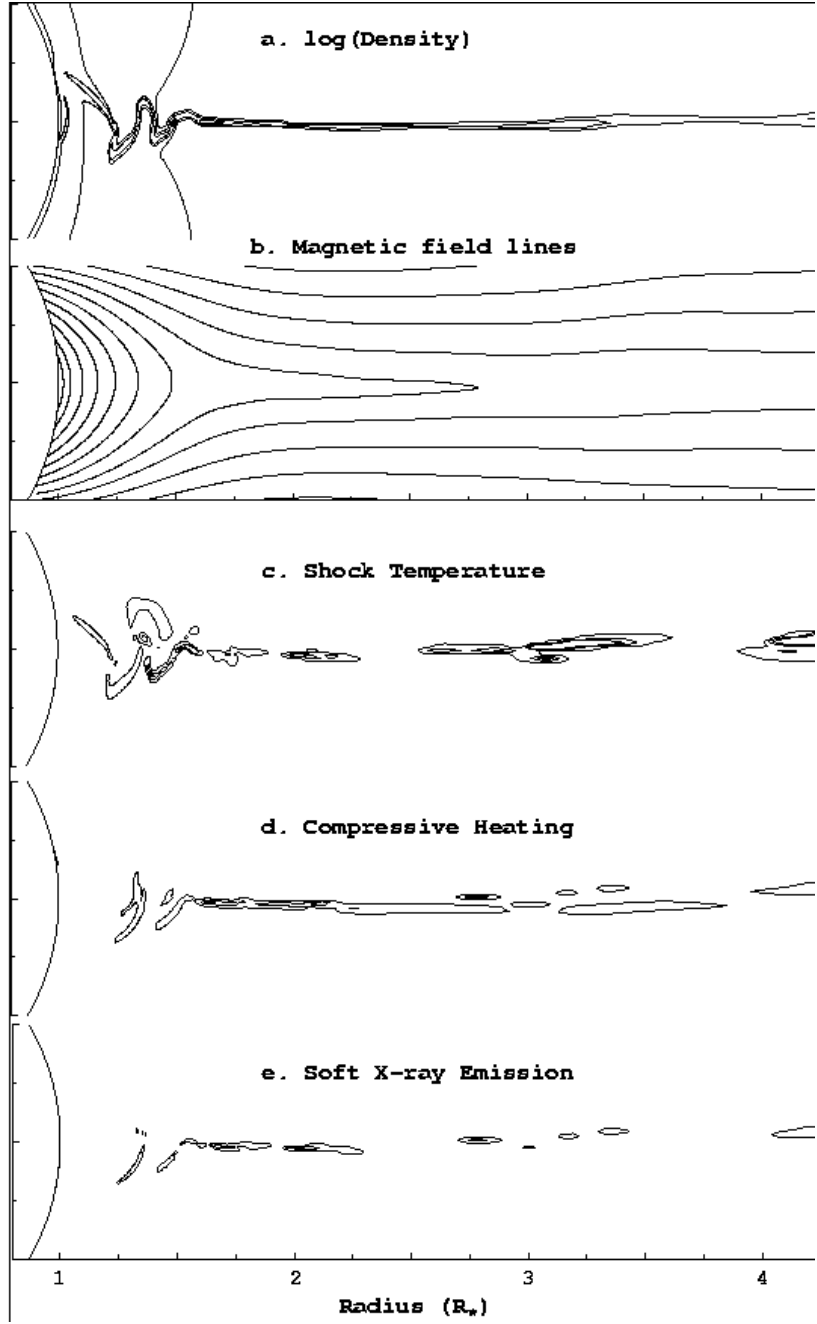}
\end{center}
\caption
{
For the strong magnetic confinement case $\eta_{\ast}=10$, contours of
a. log(density), 
b. magnetic field lines,
c. shock temperature [estimated from eqn. (\ref{tsdef})], 
d. compressive heating [computed from eqn.(\ref{qdef})],
and
e. soft X-ray emission [estimated from eqn. (\ref{qedef}),
with $E=0.1$~keV].
The 3 contour levels correspond to $\log(\rho)=$~-12, -13, 
and -14 ($g/cm^3$) in panel a; to $T_s=$~1, 11, and 21 MK in panel b;
and to $q$ (or $q_E$) of 0.15, 0.9, and 1.5
erg/cm$^3$/s in panels c and d.
}
\label{fig9}
\end{figure}

Figure \ref{fig9}c shows contours of $T_s$ computed in this way for the 
strong confinement case $\eta_{\ast}=10$.
Note that quite high temperatures, in excess of $10^7$~K, occur in
both closed loops near the surface, as well as for the open-field,
equatorial disk outflow in the outer wind.
For the closed loops, where the field forces
material into particularly strong, nearly head-on, shock collisions, 
this is as expected from previous fixed-field models
(BM97a,b).

But for the open-field, equatorial disk outflow, the high-temperature 
compression is quite unexpected.
Since the flow impingent onto the disk has a quite oblique angle, 
dissipation of just the normal component of velocity would not give a 
very strong shock compression.  
But this point of view assumes a ``free-slip'' post-shock flow, i.e., 
that the fast radial flow speed would remain unchanged by the shock.
However, our simulations show that the radial speed within the disk
is much slower.
Thus, under the more realistic assumption that incoming material 
becomes fully entrained with the disk material, i.e. follows instead
a ``no-slip'' condition, then the reduction from the fast radial wind speed
implies a strong dissipation of radial flow kinetic energy, and thus
a quite high post-shock temperature.

To estimate the associated magnitude of expected X-ray emission, we first
compute the local volume rate of compressive heating, obtained
from the negative divergence of the local kinetic energy flux,
\begin{equation}
    q \equiv - \nabla \cdot ( {\bf v} \rho v^2/2 )
    \approx - \rho {\bf v} \cdot \nabla v^2/2 \, .
\label{qdef}
\end{equation}
The contours of $q$ plotted in figure \ref{fig9}d again show that strong
compressions are concentrated toward the magnetic equator, with again
substantial levels occuring in both the inner, closed loops, as 
well as in the equatorial disk outflow. 

Let us next combine these results to estimate the X-ray emission
above some minimum threshold energy $E$, weighting the emission
by a ``Boltzmann factor'' that declines exponentially with the ratio
of this energy to the shock energy $kT_s$,
\begin{equation}
    q_E \equiv q \,  e^{-E/kT_s} \, .
\label{qedef}
\end{equation}
Using the conversion that a soft X-ray energy threshold of $E=0.1$~keV 
corresponds roughly to a temperature of $1.1 \times 10^6$~K, 
the contours in figure \ref{fig9}e show that soft X-rays above this energy
would again be produced in both the inner and outer regions of the 
equatorial disk.

Figures \ref{fig9} d and e show that the {\it volume} for flow compression and 
associated X-ray emission is quite limited, confined to narrow disk about 
the magnetic equator.
Nonetheless, the strength of this emission can be quite significant.
For example, volume integration of the regions defined in figure \ref{fig9}d 
give a total rate of energy compression
$L_c \sim 10^{36}$~erg/s, which represents about 25\% of
the total wind kinetic energy,
$L_w \sim {\dot M} v_\infty^2/2 \sim 4 \times 10^{36}$~erg/s.

This is consistent with the fraction of mass loss in the slowly
outflowing, equatorial disk, which has a value 
${\dot M_{eq}} \sim 5  \times 10^{-7} ~ M_{\odot}$/yr, 
or about the same 25\% of the total wind mass loss rate 
${\dot M} \sim 2 \times  10^{-6} \, M_{\odot}$/yr.
The terminal speed within this disk, $v_{eq} \sim$~1000~km/s,
is about a third of that in the wind, $v_{\infty} \sim$~3000~km/s,
implying nearly an order magnitude lower specific kinetic energy.
The  `missing' energy associated with this slow disk outflow thus 
represents roughly the total wind flow kinetic energy dissipated by
the flow into this slow disk.

Finally, integration of the Boltzmann-weighted emission in 
figure \ref{fig9}e gives an estimate for the soft X-ray emission above
0.1~keV of $L_x \sim 10^{35}$~erg/s.
This is substantially higher than the canonical X-ray emission 
associated with intrinsic wind instabilities, 
$L_x \sim 10^{-7} L_{bol} \sim 4 \times 10^{32}$~erg/s.
This supports the general notion that hot-stars with anomalously 
large, observed X-ray luminosities might indeed be explained by flow 
compressions associated with wind-magnetic channeling.

While this analysis thus provides a rough estimate of the X-ray 
emission properties expected from such MHD models of wind magnetic 
confinement, we again emphasize that quantitative calculations of 
expected X-ray emission levels and spectra will require a future,
explicit treatment of the wind energy balance.
 
\section{Result Summary}

We have carried out 2D MHD simulations  of the effect of stellar dipole
magnetic fields on radiatively driven stellar winds.
The initial simulations here are based on idealizations of isothermal 
flow driven outward from a non-rotating star by a strictly radial line-force.
The principal results are summarized as follows:

\begin{enumerate}
\item
The general effect of magnetic field in channeling the stellar wind
depends on the overall ratio of magnetic to flow-kinetic-energy density,
as characterized by the wind magnetic confinement parameter, $\eta_{\ast}$, 
defined here in eqn. (\ref{wmcpdef}).
For typical stellar and wind parameters of hot, luminous supergiants
like $\zeta$~Pup, moderate confinement with $\eta_{\ast} \sim 1$
requires magnetic fields of order 100~Gauss.
%
The results of standard, spherically symmetric, non-magnetic wind 
models are recovered in the limit of very small magnetic confinement,
$\eta_{\ast} \le 0.01$.

\item 
For moderately small confinement, $\eta_{\ast}=1/10$, the wind extends
the surface magnetic field into an open, nearly radial configuration.
But even at this level, the field still has a noticeable global influence
on the wind, enhancing the density and decreasing the flow speed near
the magnetic equator.

\item
For intermediate confinement, $\eta_{\ast}=1$, the fields are still 
opened by the wind outflow, but near the surface retain a 
signifcant non-radial tilt, channeling the flow toward the 
magnetic equator with a latitudinal velocity component as high as 300~km/s.

\item
For strong confinement, $\eta_{\ast}=10$, the field remains closed in
loops near the equatorial surface. Wind outflows accelerated upward from
opposite polarity footpoints are channeled near the loop tops into 
strong collision, with characteristic shock velocity jumps of up to 
about 1000~km/s,
strong enough to lead to hard ($>~1$~keV) X-ray emission. 

\item
Even for strong surface fields, the more rapid radial
decline of magnetic vs. wind-kinetic-energy density means the field 
eventually becomes dominated by the flow, and extended into an open 
configuration.

\item
The compression of open-field outflow into a dense, slowly outflowing
equatorial disk can lead to shocks that are strong enough to produce
quite hard X-ray emission, 
a possibility that 
was completely unaccounted for in previous fixed-field analyses
that focussed only on X-ray emission within closed loops
(BM97a,b).

\item
In contrast to these previous steady-state, fixed-field models, 
the time-dependent dynamical models here indicate that
stellar gravity pulls the compressed, stagnated material within closed loops
into an infall back onto the stellar surface,
often through quite complex, intrinsically variable flows that
follow magnetic channels randomly toward either the north or south loop 
footpoint.

\item
Compared with expectations of previous semi-analytic, heuristic 
analyses, the dynamical simulations here show some distinct differences 
in the overall properties of field and flow, for example with a narrower
region of equatorial confinement, and an Alfven radius that is 
lower at the pole and higher at the equator.

\item 
The outflows from magnetic poles show quite gradual non-radial area expansions, 
with terminal wind speeds enhanced by factors
(generally $ \le $1.3) that are much less than the factor three
increase predicted by previous 1D flow-tube analyses 
(MacGregor et al. 1988).
The more modest level of wind velocity modulation seen in full MHD 
simulations here is much more compatible with  observed blue-edges of 
absorption troughs in UV wind lines from hot stars.

\item 
Finally, these MHD simulations have many properties relevant to interpreting
various kinds of observational signatures of wind variability and
structure, e.g. UV line discrete absorption components;
red-shifted absorption or emission features; enhanced density-square
emission; and X-ray emission.

\end{enumerate}

In the future, we plan to extend our simulations to include
non-radial line-forces, an explicit energy balance with
X-ray emission, and stellar rotation.
We then intend to apply these simulations toward
quantitive modeling of the various observational signatures of 
wind structure that might be associated with magnetic fields
in hot stars.

\def\blankline{\par\vskip \baselineskip}
\blankline
\noindent{\it Acknowledgements.}
This research was supported in part by NASA grant NAG5-3530 
and NSF grant AST-0097983 to the
Bartol Research Institute at the University of Delaware.
A. ud-Doula acknowledges support of NASA's Space Grant College
program at the University of Delaware.
We thank D. Cohen, C. deKoning, and V. Dwarkadas for helpful discussions and 
comments.

%

\end{document}